# Silicon Optical Phased Array with High-Efficiency Beam Formation over 180 Degree Field of View

Christopher T. Phare[1,2], Min Chul Shin[1], Steven A. Miller[1,2], Brian Stern[1,2], and Michal Lipson[1]*

[1]Department of Electrical Engineering, Columbia University, New York, NY 10027, USA

[2]School of Electrical and Computer Engineering, Cornell University, Ithaca, NY 14850, USA

*Corresponding Author: ml3745@columbia.edu

Chip-scale optical phased arrays[1–15] could enable compact beam steering and LIDAR for autonomous vehicles, precision robotics, and free-space optical communications. Because these applications demand wide angle beam steering as well as high optical power in the output beam, a natural design choice would be to space the array emitters at a half-wavelength pitch, as is common in radiofrequency phased arrays. Optical phased arrays, however, unlike RF phased arrays, have been limited by the tradeoff between field of view (i.e. angle steering) and beamforming efficiency (i.e. optical power in the output beam). This tradeoff exists because optical phased arrays rely on waveguides as emitters, which suffer from strong crosstalk when placed in close proximity relative to their mode size. Here we overcome these limitations and demonstrate a platform for optical phased arrays with 180° field of view, where more than 72 percent of the power is carried in a single diffraction-limited beam even when steered up to 60° off-axis. Our platform leverages high index-contrast, dispersion-engineered waveguides spaced one half-wavelength apart without incurring crosstalk.

Optical phased arrays to date, despite ever-increasing array size and element count[6,11], remain limited in beam quality and steering angle because of unwanted coupling between dielectric waveguides, which forces radiating emitters to be placed several microns apart. Emitter spacing greater than half a wavelength steals optical power from the main beam, redirecting it into unwanted periodic grating lobes tens of degrees away that limit the uniquely addressable steering range of the array. While strong confinement in metals makes half-wavelength spacing possible in RF arrays, the nearby plasmon resonance makes such approaches[16,17] impractically lossy at optical frequencies. Kossey *et al.*[12] have demonstrated half-wavelength emitter pitch in a 5-element array of dielectric waveguides without controlling crosstalk, but that approach works only for very small end-fire arrays, where waveguides run parallel for only a few microns. Creative solutions to achieve wide field of view, like non-uniform emitter pitch[4,6,18], successfully suppress grating lobes but sacrifice power in the main beam by redirecting power carried in grating lobes over many angles rather than back in the desired direction. In LIDAR systems, this redirected power causes multiple problems, both decreasing the signal to noise ratio from the desired target and raising the overall background of returned signal, obscuring far-away or low-reflectivity targets and creating false images. In these phased arrays, since an emitter pitch of 2-3 μm is required to avoid crosstalk, power in the main beam, even theoretically, rarely exceeds 30-40 percent of the total array output[18].

We simultaneously achieve wide steering angle and high beam efficiency by designing an end-fire optical phased array at true half-wavelength pitch, engineered to avoid coupling between closely-spaced output waveguides even over millimeter-scale propagation lengths. In order to minimize coupling of modes of adjacent waveguides, we minimize their overlap in phase space instead of real space[19]. This mismatch can be seen in Figure 1a, where we show the ω-k dispersion diagram for two adjacent waveguides, with widths 300 and 400 nm. At the operation wavelength the waveguide confinement causes a clear mismatch in k-space. We achieve this k-mismatch while ensuring that all the waveguide widths are kept similar, critical to ensure uniform illumination across the array, by designing the waveguide cross section to have high dispersion ($\partial \beta / \partial w$, Figure 1b). The dispersion makes a small variation in width between adjacent waveguides induce a strong shift of propagation constant. Since maximum power coupling between two waveguides varies as

$1/((\Delta\beta/2\kappa)^2 + 1)$ where $\Delta\beta$ is the difference in propagation constant and $\kappa$ is the field overlap coupling strength[20], this engineered phase velocity mismatch allows weak coupling between waveguides even when their evanescent fields overlap. In our design, waveguides are phase-mismatched with both their nearest neighbor and with their second-nearest neighbor (separated by $\lambda/2$ and $\lambda$, respectively) by cycling through a set of three widths (300, 350, and 400 nm, in sequence). Third-nearest neighbors, separated by $3\lambda/2$, have equal widths. The 250 nm waveguide height ensures all of these waveguides are single-mode for TE polarized light. We find that even for several-mm propagation lengths, coupling between nearest, second-nearest, and third-nearest neighbors is below 17 dB and power propagates only in the waveguide into which light was launched, as shown in Figure 1d.

We demonstrate wide-angle beam steering with a 64-channel thermooptically-steered silicon phased array. Figure 2a shows a schematic of the chip design. A tree of cascaded 1x2 MMI splitters distributes the power of a 1550 nm input laser into 64 separate output waveguides. Each of these waveguides acts as a phase shifter when a thin platinum wire, 50 nm thick by 700 nm wide and 300 μm long, is placed on top of the 1 μm cladding oxide. These microheaters are designed to induce highly localized heat and minimize thermal crosstalk. Figure 2b shows a finite element method simulation of the thermal phase shifter, demonstrating a temperature change localized to within a few micrometers of the waveguide. Phase shifters are placed on a 20 μm pitch to further minimize thermal crosstalk; an optical micrograph of the phase shifters and splitter tree is shown in Figure 2c. After the microheaters, we route the waveguides in a nested 90° bend to form the final 775 nm output pitch. The scanning electron micrograph in Figure 2d shows the final varying waveguide widths on equal pitch. A 200 μm bend radius is sufficiently large to not significantly perturb the effective index of the waveguides and thus minimizes coupling as they are brought close together. Waveguides run parallel for up to 680 μm and then terminate in a low-roughness plasma-etched facet[21]. At this facet, the waveguides act as emitters with a quasi-dipole emission pattern. To control the large number of phase shifters, we use thick aluminum wires to route each microheater to a bond pad at the chip perimeter and connect the opposite terminal of the heaters together in a low-resistance common ground. We mount the chip with high thermal conductivity silver sintering paste directly to an aluminum heatsink and then wirebond each heater to an attached printed circuit board. Figure 2e shows the packaged chip. A 64 output digital to analog converter drives each microheater individually, allowing arbitrary phase control over more than 2π.

We experimentally achieve grating-lobe-free operation over an entire 180° field of view. Figure 3a shows emitted far-field beam profiles with the beam steered up to ±80° off-axis. In order to compensate phase mismatch in each of the 64 channels when no heater is applied, we place a single-element, 1 mm diameter photodiode in the Fraunhofer far-field approximately 20 cm away from the chip and use a global optimization algorithm (see Supplementary Information) to align the phases and form a beam centered on the fixed detector position. To measure emission over such a large field of view while simultaneously avoiding optical aberrations from a high numerical aperture measurement system, we design a mechanically scanned far-field imager by placing the photodiode on an optical rail and rotating it in the waveguide plane about the chip output facet. Total phase shifter power consumption regardless of beam position is approximately 600 mW.

Our array emits a high quality beam with 11.4 dB peak to sidelobe power ratio, close to the theoretical maximum of 13 dB (fundamentally limited by the sinc$^2$ diffraction pattern of a uniformly-illuminated rectangular aperture). This peak to sidelobe ratio is achieved over the entire 180° field of view, and it is to our knowledge the highest demonstrated to date. Figure 3b shows one beam in logarithmic scale to make sidelobe levels easier to visualize. Even when the beam is steered 70° off-axis, the peak to sidelobe power ratio drops only slightly, to 11.0 dB, simply due to the fundamental cos$^2$ dipole-like emission pattern of a single waveguide emitter. The beam width (FWHM) is limited by the diffraction from the 49.6 μm wide aperture to 1.6°. When steered to sharp angles, the beam broadens (to 2.9° at 60° off-axis) because the glancing angle decreases the apparent size of the aperture. Residual sidelobe levels above the sinc$^2$ pattern are limited only by the uncontrolled phase from two malfunctioning microheaters and by phase uncertainty in each channel caused by detector noise in the beam convergence process.

The array shows at least twice the optical efficiency expected in randomized coarse-pitch arrays, concentrating 72% of the total emitted power in the main 0° beam. Theoretically, with perfect phase alignment, our tightly-packed waveguide approach would direct a full 90% of the total output power into the main beam (corresponding to the integral of a sinc$^2$ function between the first nulls), compared to a theoretically expected 30-40 % in randomized coarse-pitch arrays.

We have demonstrated an all-dielectric optical phased array with true half-wavelength emitter pitch that generates nearly diffraction-limited output beams over a full 180° field of view. Our anti-crosstalk design maintains beam quality even at very small emitter spacing, allowing high fill factors, and scales to a large number of channels. Such systems will enable the next generation of extremely compact and lightweight long-range LIDAR.

**Acknowledgements**

Funding was provided by the Defense Advanced Research Projects Agency (DARPA) (HR0011-16-C-0107). This work was performed in part at the City University of New York Advanced Science Research Center NanoFabrication Facility.

**Author Contributions**

C.T.P., B.S., and M.L. conceived the work. C.T.P. designed and fabricated the devices, conceived and implemented device testing, and wrote the manuscript. M.C.S. designed and carried out chip packaging. S.A.M. assisted with testing. S.A.M. and B.S. wrote the beam convergence algorithm. M.L. supervised the project and edited the manuscript.

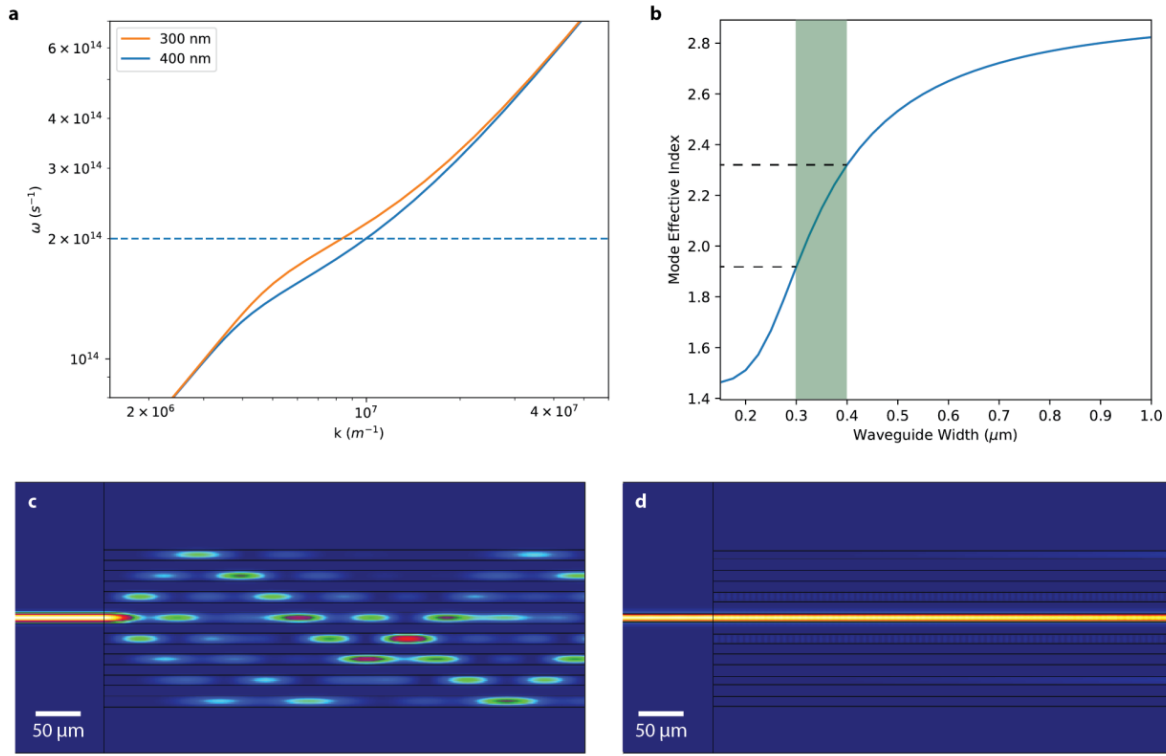

**Figure 1: Waveguide array avoided coupling. a.** Omega-k dispersion relation for two widths of 250 nm thick Si waveguides. At the operation frequency, adjacent waveguides are well-separated in k space. **b.** Effective index versus waveguide width. We choose widths between 300 and 400 nm (green shading) to take advantage of the steep index slope. Keeping waveguide widths roughly equal but well-separated in index is key to creating a uniformly-illuminated aperture. **c.** Eigenmode expansion simulation of an array of eight waveguides with equal width of 400 nm on 775 nm pitch. Light is launched into a single waveguide from the left and couples strongly when adjacent waveguides are introduced. **d.** Simulation of the waveguide array with sequentially varying widths of 300, 350, and 400 nm. Light is launched into the 300 nm wide waveguide, and less than 3% coupling is observed, only to the waveguides with equal width.

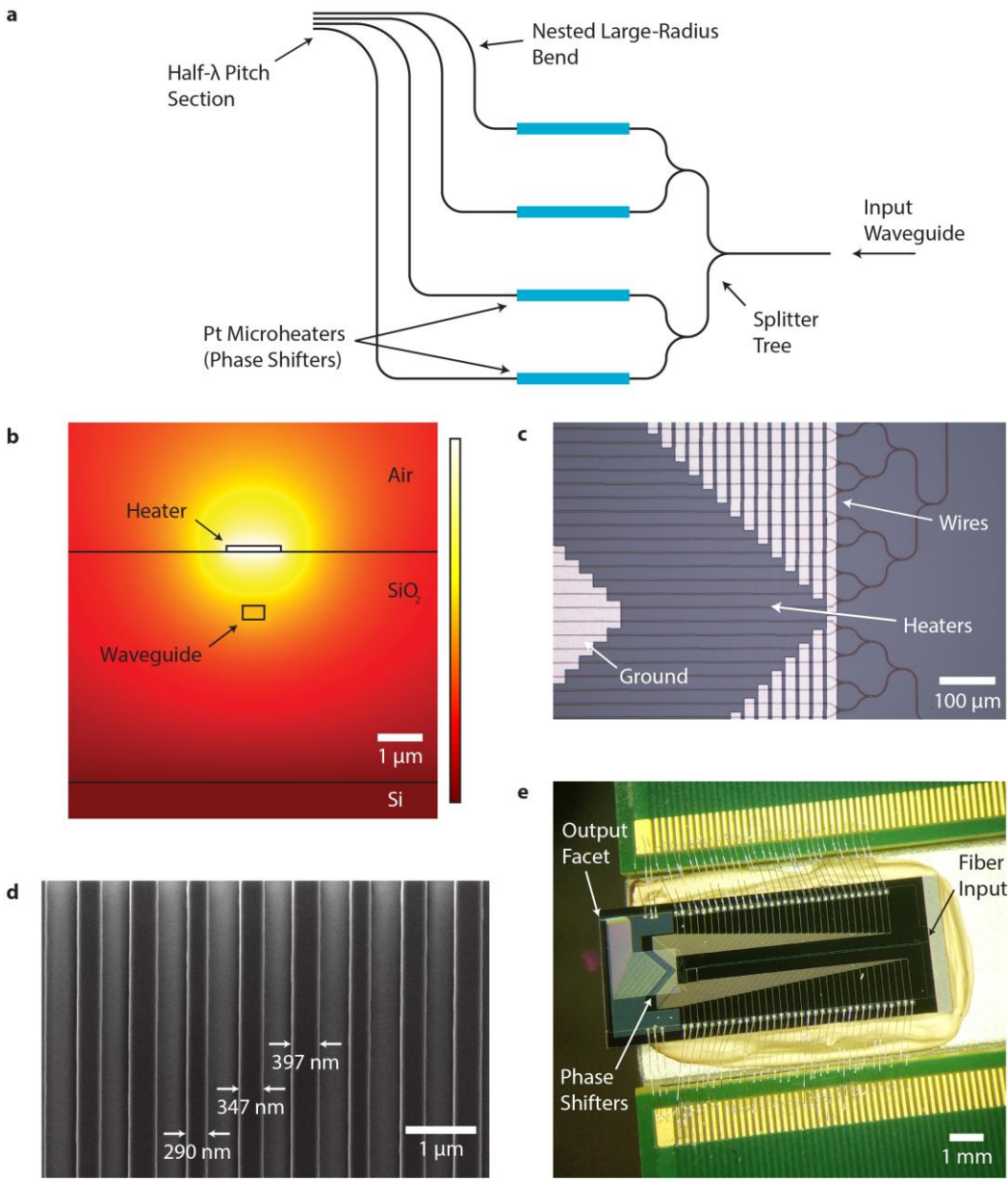

**Figure 2: Fabricated device. a.** Schematic of chip design. Light enters at the right and travels sequentially through the splitter tree, thermal phase shifters, and the narrow-pitch output section. **b.** Finite element method simulation of thermal phase shifter. Heat is localized to within a few micrometers of the heater and waveguide. **c**. Optical micrograph of part of the splitter tree, thermo-optic phase shifters, and aluminum wires. **d.** Top-view scanning electron micrograph of the varying waveguide widths on equal 775 nm pitch. **e.** Low-magnification optical image of the wire-bonded chip on its heatsink. Over 90% of the chip area is dedicated to fan-out wiring to match the 200 μm trace pitch on the PCB. Note the output facet overhangs the heatsink to ensure an unobstructed 180° view of the output beam.

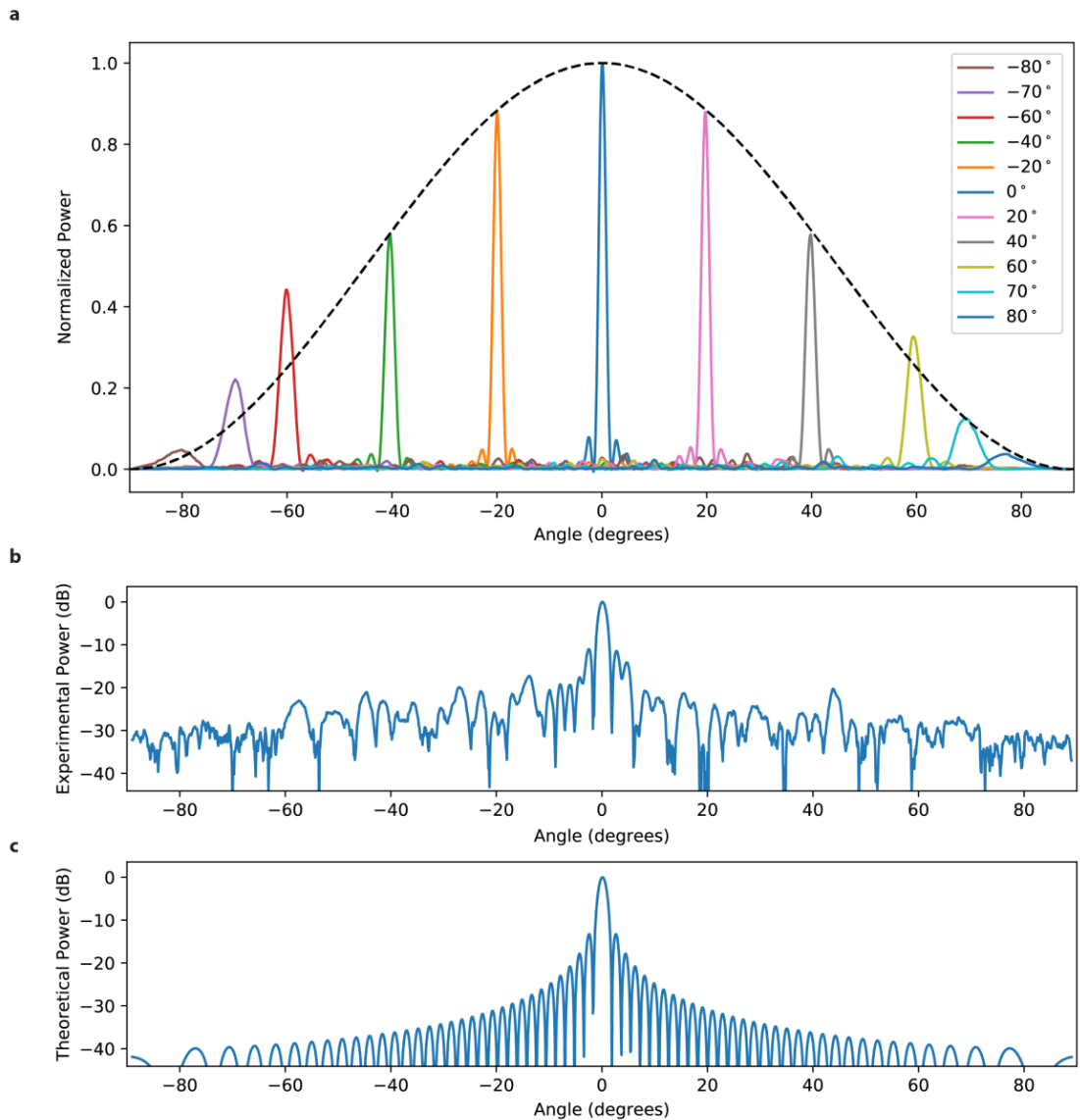

**Figure 3: Far-field beam measurements. a**. Measured far-field optical power versus output angle over a 180° field of view as a function of target beam angle, for target angles up to ±80°. Power is normalized to the 0° peak, but relative amplitudes are as-measured. The dashed line corresponds to the expected attenuation from the dipole-like emitter envelope. **b.** Logarithmic scale plot of beam steered normal to the array output. Peak to sidelobe ratio is 11.4 dB, with all but the nearest sidelobes below 17 dB. **c.** Sinc[2] theoretical emission profile fit to data in (b) with a single aperture width parameter.

**Supplementary Information**

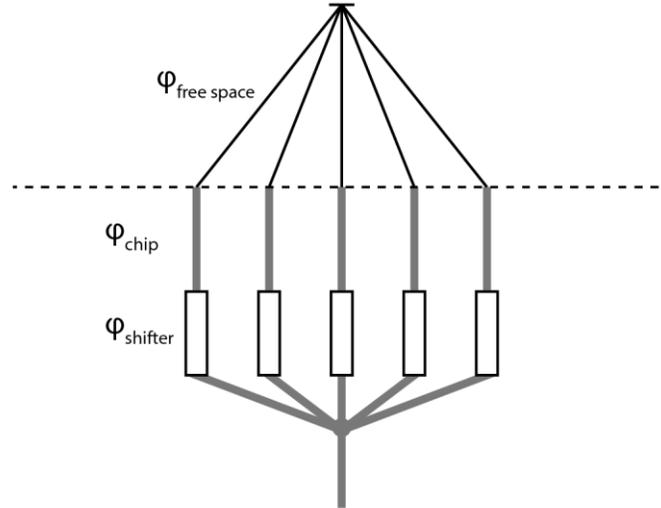

**Supplementary Figure 1:** Conceptual phased array for convergence algorithm. The entire system from laser to detector (including free-space portions) is a multichannel interferometer, with phase contributions from deliberate phase shifting as well as on-chip and free-space propagation length.

Consider the model N-channel phased array in Supplementary Figure 1. After exiting the 1:N splitter, each channel acquires a phase offset from multiple contributions: $\phi_{shifter}$ from the deliberate controllable phase shifter, $\phi_{chip}$ from propagation length between the splitter and the output facet, and $\phi_{free\text{-}space}$ from the distance between the output facet and the target or detector. We the represent the total phase of the light from channel $n$ at the detector as $\phi_n = \phi_{shifter} + \phi_{chip} + \phi_{free\text{-}space}$. In the far-field in broadside (0°) operation, all the lengths $\phi_{free\text{-}space}$ from each channel to the detector are equal, so forming a beam at the detector is equivalent to lining up the phases at the output facet of the chip. Similarly, moving the detector left or right in the far field lengthens some $\phi_{free\text{-}space}$ with respect to others, requiring a phase tilt at the facet for constructive interference at the detector. Bringing the detector closer, into the Fresnel region, makes the outer channels travel noticeably farther than the inner channels, so constructive interference requires a phase curvature

at the chip facet and therefore a focusing beam. In all cases, the beams from each channel coherently sum:

$$E_{det} = \sum_n a_n e^{i\phi_n}$$

where $a_n$ is the amplitude of channel n, and our goal is to choose the set of $\phi_n$ such that $|E_{det}|^2$ is maximized. We find that set of $\phi_n$ through a genetic algorithm-based global optimization that accounts for channel non-ideality and crosstalk (thermal and electrical) between phase shifter channels. A genetic algorithm (GA) is a procedure for optimizing functions of many variables, functioning analogously to biological evolution[S1]. In order to find an optimized set of $\phi_n$, we use a GA with $|E_{det}|^2$ as the fitness function to be maximized, as measured by a photodetector at the desired beam position. In our GA procedure, a population of sets of $\phi_n$ values is first initialized to random values corresponding to voltages for the phase shifters. In each generation, or iteration, the sets in the population are evaluated, and new sets are created from combinations of parent sets in a crossover process. Sets having higher fitness values more likely to pass on $\phi_n$ values to the following generation. In addition to crossover, mutation events occur in each generation, whereby $\phi_n$ values are randomly varied. By this procedure, successive generations approach an optimized set of $\phi_n$ values and the beam is converged onto the detector's position.